\documentclass[a4paper,man,natbib]{apa6}

\usepackage[english]{babel}
\usepackage[utf8x]{inputenc}
\usepackage{amsmath}
\usepackage{graphicx}
\usepackage{caption}
\usepackage{color}
\usepackage{makecell}

\usepackage[english]{babel}
\usepackage{amsfonts}
\usepackage[colorinlistoftodos]{todonotes}
\usepackage{algorithm}
\usepackage{algpseudocode}
\usepackage{geometry}
\usepackage[colorinlistoftodos]{todonotes}

\usepackage{amsthm}
\newtheorem{definition}{Definition}
\usepackage{float}
\title{Estimation of Time Delay caused by Point Geometry in Public Transport}
\shorttitle{Estimation of Time Delay}
\author{\textsuperscript{1*}Muhammad~Naeem, \textsuperscript{1}Mehdi~Katranji, \textsuperscript{1}Guilhem~Sanmarty, \textsuperscript{1}Sami~Kraiem, \textsuperscript{2}Hamza~Mahdi~Zargayouna and \textsuperscript{1}Fouad~Hadj~Selem}
\affiliation{- \\\textsuperscript{1}VEDECOM Institute \\ 77 Rue de chantiers, 78000 Versailles \\ France
	\\-\\
	\textsuperscript{2}IFSTTAR \\
	14 Boulevard Newton, 77420 Champs-sur-Marne \\ France
\\-\\Email: \\ 
\textsuperscript{1}FirstName.LastName@vedecom.fr\\
\textsuperscript{2}hamza-mahdi.zargayouna@ifsttar.fr\\
*corresponding author: Muhammad.Naeem@vedecom.fr}

\abstract{Travel time prediction is a well-renowned topic of research. It is primarily influenced by traffic congestion, road conditions and route geometry. Among them route geometry at any point is not investigated enough to find a sound spatial dynamics of timing delays and quality of public transport. This study investigates the reliability of travel time to build a new key performance indicator of public transport network. We have introduced a suitable point wise clustering followed by an adapted statistical significance analysis. The outcome is an estimation of absolute delay at geometrical points independent of the delay at bus station. This outcome serves as an incremental delay towards overall delay. Our investigation suggests that this novel metric of delay time contributes around 23.56\% in overall delay. The proposed methodology (here in known as Delay Time and Quality of Service ($DTQS$)) is capable of providing novel insights which attracts attention among policy makers of public transport operators. The novel insight is coined with assignment of high value to level-of-service attributes. Moreover, the proposed technique is independent of any assumption of normal distribution.}


\begin{document}
\maketitle

\section{Introduction}
{P}{ublic} transport is a part and parcel of urban life. Improvement in the quality of public transport plays an essential role in the quality of urban mobility. It goes without saying that the prevention control in delays in the public transport stays at the heart of quality control metrics. It implicitly as well as explicitly governs any system which can monitor the parameters of public transport quality. A prime parameter useful to observe the bus lines functionality is the accurate estimation of time delay. It assists in maintaining an adequate service quality with a reasonable certainty. Low reliability, slow speed and increase load of passengers accumulate towards a situation which ends up in a critical situation for the bus operator \cite{Ryus2003-ny}. Finding the delay or early arrival at a bus station may not be very complicated with the advent of the technological era. The continuous monitoring for over a long period of time enables to provide an immediate analysis to transport operators around pros and cons of delay or early arrival at a specific bus station. However, public transport operators are interested beyond this point. There is a motivation triggered by the needs and wants to find the implicit root cause for a delay at a single or many bus stations other than the obvious reasons at designated bus station. It was reported that on average 60\% of travel time in public transport in Beijing China is claimed by delays at intersection \cite{zhang2011computing} and \cite{liu2013intersection}. It is an interesting problem to find the correlation lurking between the point geometry and absolute delay status at that point independent of its previous delay (or relative delay). This interesting problem turns into a challenge when the bus data is insufficient to explicitly calculate the relationship between point geometry and timing given a certain number of conditions such as busy or idle hours. 
\\
The sophisticated devices such as global positioning system (GPS) receivers provides an opportunity to draw out a travel route profile for a specific vehicle. The vehicle profile is able to estimate the intersection delays whether they are running time delays, acceleration, deceleration or stop delays. For a single vehicle, this is possible merely by examining its route profile. However, on single route, different vehicles face different situations over a range of various time slots. Hence, hardly a single route is not sufficient to provide strong estimation as it is not consistent enough to consider for a long period of time. The term delay is too broad in context of formulating the bus profile based on the fact that there are numerous types of delays. The delays are caused by traffic congestion, passenger boarding, bus station stop, road delays resulted by specific geometry of the route. Among of these, we have chosen the road delays. \cite{liu2013intersection} emphasizes the importance of road delays specially on intersections. We have made it generalize that the road delays are not limited to intersection delays rather it can happen at any other point as well. Hence we have introduced the term point delay. We will illustrate this thanks to data provided by local transport company on their bus network in {\^I}le-de-France as shown in the figure \ref{fig:rejeu.course} and \ref{fig:rejeu.loc}. To achieve the major goal of analysis on point delay, we face following challenges to fix to provide a timing quality metrics. These challenges are as below:
\subsection{Integration problem}
It is a common practice that the transport operators collect data using different business process and services. Occasionally, these services are incoherent with each other suffering from the problem of mismatching especially if the data collection is made by two distinct third-parties like in our case. At individual scale, such data can be exploited with certain conditions and narrowed usability. However, it is required to merge them well in order to perform a knowledge discovery operation using entire set of information instead of incoherent chunks of datasets. This problem was evident in our case where we obtain two different datasets which we shall discuss in forthcoming section.  
\subsection{Stop time exclusion}
In computation of the overall delay, there are many of its components attached to the profile of a vehicle. Techniques have been proposed for initial deceleration delay, queue move time, stop delay followed by final acceleration delay. The same has been discussed with proposal of different models as in \cite{Roess1997-zg} and \cite{Ko2008-py}. This aspect directly influence the further delay on route. So we must have to exclude the stop time delay ($std$) in all of its possible shapes.
\subsection{Dead mileage}
This is the out of service transit period. It is a phenomenon known when the bus completes its route and is ready to start a new route. We shall term it as transit period short ($tps$). In situations when the bus is equipped with GPS devices recording at the a range of specific frequency, the bus will continue on receiving the GPS coordinates during this time. There is also a situation when the bus goes on off service. For such situation know as deadheading, the bus usually moves towards its depot. We term it as transit period long ($tpl$). A mechanism is required to mark both of these transit period separately.
\subsection{Point delay}
Apart from the delays mentioned above, there is another type of delay which is related to operational delay between two bus stations. This is point delay $pd$ which is more complicated to find. It has its root in stipulated speed at specific road segment and designated limits of speed profile as leveraged by the bus operator. In fact, when the bus is in its delay, this might be due to traffic congestion or complexity of the route (narrow and tedious path). Usually if this happens a few times then it is ignorable. However, there is a need to point out the significance level of delay at particular point(s). 
\subsection{Reconstruction of route} 
Reconstruction of correct route from incomplete data is also a challenge when there is an uneven jump from one station to other station skipping intermediate stations. The precision of the GPS device is important because as the precision is less the GPS coordinates in records are a bit out of the route. The points can be shown on a park or a parallel road or street. Such GPS coordinates not only to be identified but also corrected in pursuit of reconstruction of the route. 
\par
Keeping in view of the problems mentioned above, we examined the research work introduced by other researcher. We noticed that more research has been carried out for various types of delays except pd. Our objective in this study is to devise a technique such that we can exactly find the pd. We in this study have proposed a tool to assist the transport operator to find the relations between time status and geometry defined by geographic coordinates. The underlying methodology (herein named as Delay Time and Quality of Service ($DTQS$)) of the tool has been validated over bus data provided by local transport company in one sector of {\^I}le-de-France. $DTQS$ reflects the degree of variation in the delay time of a trajectory that is repeated in similar conditions across a period of three months. $DTQS$ can play a key factor that bus operator put in consideration while drawing out basic trajectory decisions, such as decisions regarding mode, route and departure time. Furthermore, it has a direct impact on travel time variability which is significantly discussed in literature \cite{Duran-Hormazabal2016-in}. They also shown statistically that public buses running in mixed traffic are suffering more to traffic perturbation. It means specific geometric points are more important to be known to transport operator if the bus has to run outside its marked lanes. Our proposed methodology helps transport operators in their attempts to quantify how much people value reductions in their travel time variability. 
\par
The prime contribution of our study is the technique of estimation of point delay irrespective of stop delay, turn delay etc. We focus on the spatial dynamics of the geometric parameters of the road. Researchers pointed out that the intersection delays account for 17\% to 35\% of the total travel time \cite{nielsen1998using}. However,this range is quite broader and needs to be more specific. The proposed geometry point delay is more generalized to cover the intersection delay as well. Moreover, our finding that the point delay accumulates around 23.5\% of overall delay is more specific.
\par
The rest of the paper is divided into five sections. In next section, we shall examine the central topic of this research in context of work carried out by other researchers. The section of methodology itself is composed of many subsections explaining our technique for how we achieve our goal. In empirical evidence, we discuss the maps resulted as an outcome of application of the methodology. In the last section, we conclude with brief summary of this study.
%%%%%%%%%%%%%%%%%%%%%%%%%%%%%%%%%%%%%%%%%%%%%%%%%%%%%%%%%%%%%%%%%%%%%%%

\section{Literature Review}
One of the earlier work was carried out by \cite{nielsen1998using} in estimation of delay at a specific point in the route. Their work was based on survey data in Copenhagen, Denmark. They worked on intersection turn delays in medium congested road network of the city of Denmark. The outcome of their research study concluded in a wide range of time delays from 17\% to 35\% with focus on turn delays only. However, intersection turns are not the only geospatial points where calculation of turn delay is interesting. A broader approach needs to be put forward with the concept of point delay covering the intersection points as well. However, even the intersection delays were not very much appreciated for research purpose until a classification work was proposed by \cite{zhang2011computing} in large metropolitan area of Beijing China. Intersection delays can be translated into turn costs which constitute a significant part of overall time delays. However, if we generalize this concept then its importance becomes more non trivial. Another limitation of the work carried out by \cite{nielsen1998using} was that equally representative data from survey or field experiment is usually hard to obtain and update frequently. 
\par
Later on, another work was forwarded by Ko et. al., \cite{Ko2008-py} pertinent to the domain of this study. They proposed the methodology of formulation of vehicle profile with the objective of finding the variation in speed at stop, acceleration and deceleration moment.  However, with the advent of modern technological tools, finding and maintaining delays at bus stop is quite common in modern urban transport operators. Now transport operators are more interested in one step ahead where these parameters are required in terms of specific geometry so they can put forward for the solution in considering and investigating the specific geometry at any point. It is reported that the techniques used for modeling the time delays in transports sector have been broadly clustered into three groups \cite{Choudhary2016-ks}. These models are based on machine learning \cite{liu2013intersection}, \cite{zhang2011computing}, statistical models \cite{Aron2014-od}, \cite{Duran-Hormazabal2016-in} and historical data based models \cite{Roess1997-zg}, \cite{nielsen1998using} and \cite{Ko2008-py}. In general these models are aimed at forecasting the delay at a specific station during the trajectory. However, it is also required that root cause of delay be focused as well. The cause of this delay is usually characterized by congestion of traffic, passenger facilities at bus stations. By arguing, we propose the view that geometry of the path is also non trivial to find whether a delay is in norm at a specific GPS coordinates other than the designated station. Such a marking can assist the bus operator in further investigation of possible reasons of unusual delay at that point or resulting a perpetual delay. In the research of time delay, the reasons of delay have been investigated by many other aspects as well such as finding the nature of distribution of delay time. An interesting work was proposed by \cite{Aron2014-od}. They performed a comparison of six statistical distributions to model travel time in perspective of reliability. The six distributions include Lognormal, Gamma, Burr, Weibull, Normal Mixture and Gamma Mixture. They conclude that Gamma mixture and Normal Mixture distributions both correspond to the observed distribution for the fluid conditions. However both were not satisfactory in the traffic congestion time periods. In our conclusion to the existential  research, there is a strong impetus for investing the point delay as one of the root cause of the overall delay. Keeping in view of this context, we shall propose and discuss the methodology in next sections. One interesting work using the classical machine learning technique was proposed by \cite{zhang2011computing} and \cite{liu2013intersection}. They focused on prediction of turn delays. They extract the turn delays from road intersection with emphasis on parameters including angle, geographic neighborhood effect and topological relationship of road intersections. They predict the turn delay for the other intersections using the classical neural network. However, we argue that the turn delays needs to be broaden when it comes to those points which are not limited to only road intersection. The delay can occur at points other than turn delays. Conclusively, we put forward the idea that the estimation of point delay in literature is still tangential and needs to be investigated in much depth in order to find its specific and precise impact on overall delay. Keeping in view of these limitations, we have proposed our methodology to overcome the aforementioned issues in the following section.
%%%%%%%%%%%%%%%%%%%%%%%%%%%%%%%%%%%%%%%%%%%%%%%%%%%%%%%%%%%%%%%%%%%%%%%
\section{Methodology} 
There are different parameters when one need to define the bus profile with a specific goal in mind. While defining the methodology, we have trajectory, set of gps coordinates, speed profile and time delay of the bus. We define each of the them and then come up with the objective function with constraint around which we have drawn out the technique.
\begin{definition}[Collection of GPS]
It is a set of gps coordinates. Each gps coordinate is composed of latitude, longitude and in addition a time stamp with this pair. We denote this set as below:
\begin{equation}\label{eq:p}
p = \left \{ ps_{11},......,ps_{mn} \right \}
\end{equation}
$p$ is the set of $n$ points starting from first station to last m\textsuperscript{th} station. Each $ps$ is defined as:
\begin{equation}\label{eq:ps}
ps = \left \{ lat_{i}, lon_{i}, time_{i} \right \}
\end{equation} 
The $lat_{i}$ means the latitude of gps point, $lon_{i}$ means longitude of point at i \textsuperscript{th} time stamp. From equation \ref{eq:p} and \ref{eq:ps}, we can derive the definition of trajectory of the bus such as that it is sequential list of the points:
\begin{equation}\label{eq:trajectory}
tr_{j} \leftarrow p_{j}
\end{equation}
where j \textsubscript{th} set of coordinates realize the j \textsuperscript{th} trajectory. 
\end{definition}
%%%%%%%%%%%%%%%%%%%%%%%%%%%%%%%%%%%%%%%%%%%%%%%%%%%%%%%%%%%%%%%%%%%%%%%%%%%%%%
\begin{definition}[Time delays]
	The total delay $d$ during the course of the trajectory of the bus is composed of different types of delays including stop time delays ($std$), out of service transit period such as transit period short ($tps$), transit period long ($tpl$) and point delay ($pd$). Mathematically, we shall express it as: 
	\begin{equation}\label{eq:d}
	d = \sum_{1}^{i}std + \sum_{1}^{j}tps + \sum_{1}^{k}tpl + \sum_{1}^{l}pd
	\end{equation}
\end{definition}
%%%%%%%%%%%%%%%%%%%%%%%%%%%%%%%%%%%%%%%%%%%%%%%%%%%%%%%%%%%%%%%%%%%%%%%%%%%%%%
\begin{definition}[Constraints]
Let we denote speed profile by the letter $s$, a bus is subjected to follow up and defined as below. 
\begin{equation}\label{eq:s}
s = f(sp_{g},sp_{c},sp_{a})
\end{equation}
The speed profile $s$ is a function of two types of speeds; road speed limit set by government such as road authority ($sp_{g}$) and road speed limit set by transport corporation ($sp_{c}$). Certainly, it is assumed that the later ($sp_{c}$) is assumed to be less or equal to the prior speed ($sp_{g}$). However, the actual speed profile ($sp_{a}$) is more important as this is what a bus is observed with. It is required to be set under both of the other profiles albeit this profile is only affecting the point delay ($pd$). We can define the constraint based on this speed profile as below:
\begin{equation}\label{eq.constraint}
sp_{a} =
\left\{\begin{matrix}
sp_{g} & if & sp_{g} \leq sp_{c}\\ 
sp_{c}&  & otherwise 
\end{matrix}\right.
\end{equation}
\end{definition}
%%%%%%%%%%%%%%%%%%%%%%%%%%%%%%%%%%%%%%%%%%%%%%%%%%%%%%%%%%%%%%%%%%%%%%%%%%%%%%
\begin{definition}[Objective function]
The goal of this research is to find out the point delay $pd$ which can occur at any point in the course of j\textsuperscript{th} trajectory $tr_{j}$. So for the optimized performance of the buses, we need to minimize all possible values of $pd$ with following objective function.
\begin{equation}\label{eq:objective.function}
z= min
\left\{\begin{matrix}
\sum_{1}^{l}pd & -  & \sum_{1}^{i}std  & -  & \sum_{1}^{j}tps & - & \sum_{1}^{k}tpl
\end{matrix}\right.
\end{equation}
\end{definition}
%%%%%%%%%%%%%%%%%%%%%%%%%%%%%%%%%%%%%%%%%%%%%%%%%%%%%%%%%%%%%%%%%%%%%%%%%%%%%%
\subsection{Statement of the problem}
Given a bus trajectory along with its start and end point configuration under particular timing schedule over many days. There are $n$ number of points in between the two points such that a route profile can be defined as a function of set of points $p$, various time delay $d$ and speed profile $s$. We can describe it by equation.
\begin{equation}\label{eq:profile}
E = f(p,d,s)
\end{equation}
Where $E$ is route profile. The objective function in equation \ref{eq:objective.function} is to be applied on the route bus profile such that the new profile $z = f(E)$ is evolved under the constraints as laid out in the equations \ref{eq:profile}, \ref{eq.constraint} and \ref{eq:s}. For the sake of simplicity we shall name this new profile as $z$. In the new profile $z$, we are required to consider only the point delay ($pd$), the intermediate points between any two stations and the actual speed profile ($sp_{a}$). The letter $z$ provides the overall estimation of the route profile in terms of delays at specific points while respecting the above constraints. Keeping in view of the statement of the problem, we have devised an algorithm whose steps are explained in the following sub section.

\subsection{Geometry based delay finder algorithm}
\begin{figure}[H]
	\centering
	\includegraphics[width=0.489\textwidth]{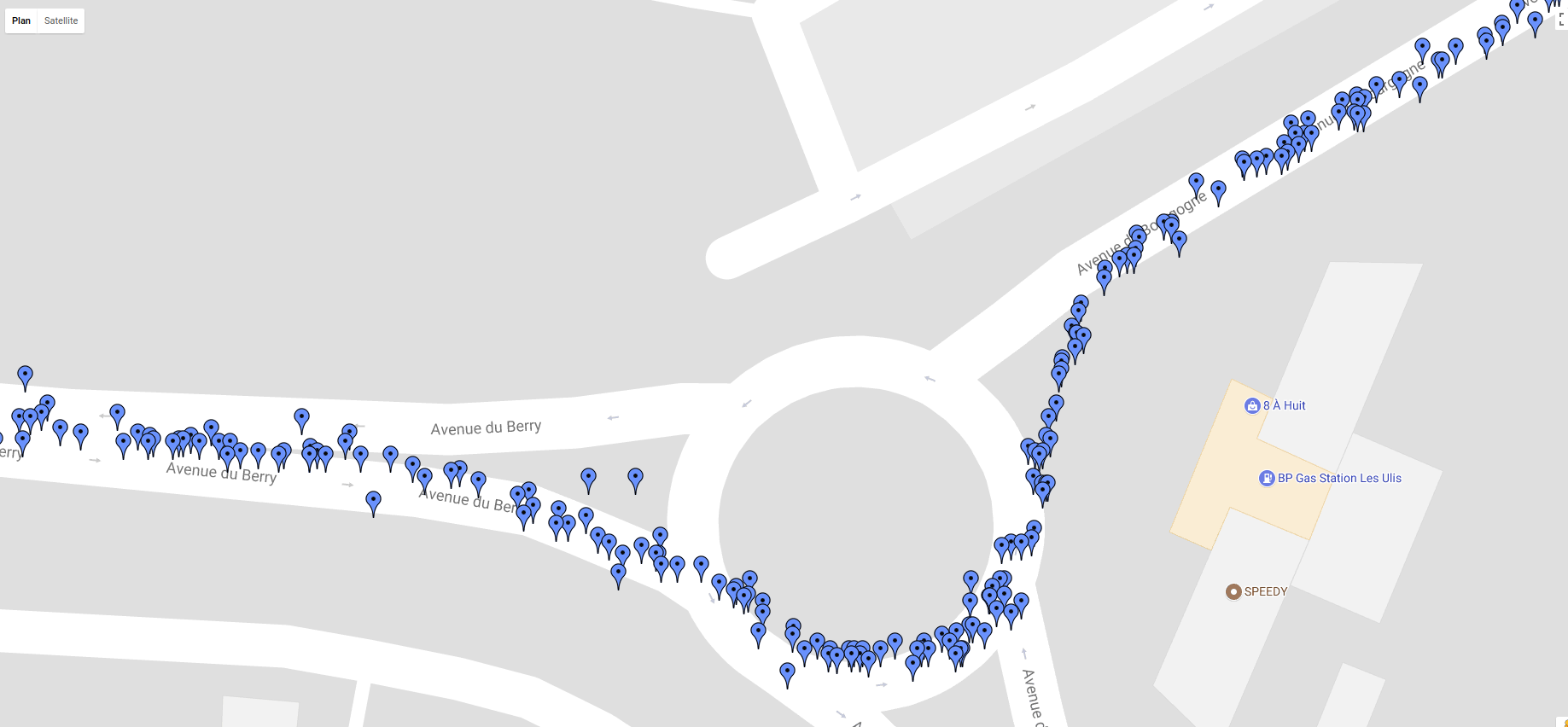}
	\caption{Sorting and aligning of the GPS coordinates. (Before)}
	\label{fig:sorting.alignment.before}
\end{figure}

\begin{figure*}
	\includegraphics[width=\textwidth]{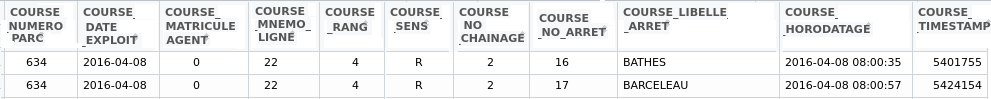}
	\caption{Rejeu course: A trajectory of a bus covering all of the stations}
	\label{fig:rejeu.course}
\end{figure*}

\begin{figure*}
	\includegraphics[width=\textwidth]{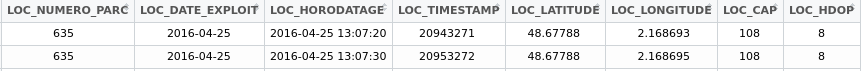}
	\caption{Rejeu location: Trajectory of the bust showing the GPS coordinates as a function of time}
	\label{fig:rejeu.loc}
\end{figure*}

Given the data, the first step is the pre processing of the data which includes cleansing of data and make it suitable for the application of the sepcific algorithms. In this step, from the pool of dataset, we identify each line of the bus given its specific parameters like sub line, chain, range, date, time etc. Then we find the gps coordinates between every two stations by mapping the first dataset to the second dataset (see figure \ref{fig:rejeu.course} and \ref{fig:rejeu.loc}). 
The second algorithm \ref{alg:mapping} identifies all of the points which are actually far away from the road as shown in the figure \ref{fig:sorting.alignment.before}. The unwanted gps points are either mapped to the correct nearest coordinate or shredded away if far away from the correct point. Here, we reorder all of the points in their true gps order. This ordering is calculable by means of parameters such as the direction of the bus, the time the bus left the gps coordinate.  
%%%%%%%%%%%%%%%%%%%%%%%%%%%%%%%%%%%%%%%%%%%%%%%%%%%%%%%%%%%%%%%%%%%%%%%%%%%%%%%%%%%
\begin{algorithm}
	\caption{Mapping of imprecise gps coordinates to bus routes}
	\label{alg:mapping}
	\begin{algorithmic}[1]
		\Function{gps\_mapping}{$tracks, ox $}\Comment{where tracks - collection of bus tracks, ox - reference network graph}
		\State $ox = build\_osmnx\_graph()$ \label{op:ox.1}
		\State $routes = load\_all\_bus\_routes()$
		
		\For{$r$ in $routes$}
				\State $gps[] = extract\_gps(r)$
				
				\State $p[] = \phi $ \Comment{array of gps points for every new route}		
				
				\For{$g$ in $gps$} 
					\State $g_n$ = $find\_nearest\_ox\_point(r,g,ox)$ 
					
					\State $address = verify\_by\_google\_map(r, g_n)$
					
					\If {$address == True$}
					\State $p = p \cup g_n$
					\EndIf
					
				\EndFor
				\State $g_n=detect\_surrounding\_points(r, g_n, ox)$ \Comment{Detect forward, backward surrounding points.}
				\State $g_n = reorder(r, g_n, ox)$ \Comment{reorder all points in their true gps order}
				
				\State $r \leftarrow p $ \Comment{update gps points in route}
		
		\EndFor

		\EndFunction		
	\end{algorithmic}
\end{algorithm}
%-------------------------------------------------------

In the second algorithm \ref{alg:delay.points.identification}, the distance between any two points is calculated. This helps in finding the cumulative distance. Then Mark out the 'out of route' gps coordinates resulted in mapping. The real time is available in the dataset of the bus but theoretical time is a compulsory requirement. Here theoretical time refers to the time usually required by a bus to travel between two points. This source can be obtained from Graphhopper \cite{Karich2014-ok,Boeing2017-yq}. 
%-------------------------------------------------------
\begin{algorithm}
	\caption{delaty point identification in bus routes}
	\label{alg:delay.points.identification}
	\begin{algorithmic}[1]
		\Function{identify\_delay\_point}{$df, ox $}\Comment{where df - data frame containts collection of bus tracks, real, theoretical time, ox - reference network graph}
		\State $ox = build\_osmnx\_graph()$ \label{op:ox.2}
		\State $routes = load\_all\_bus\_routes()$
		
		\For{$r$ in $routes$} \label{op:0} 
			\State $gps[] = extract\_gps(r)$
			\State $lat[] = extract\_latitude(gps)$
			\State $lat[] = extract\_latitude(gps)$
			\State $lon[] = extract\_longitude(gps)$
			\State $time[] = extract\_time_stamp(gps)$
		
			\State $n = length(gps) $
		
			\State $d[] = t[] = \phi $ \Comment{array of consecutive distance and delay time between two gps points}		
		
			\State $d[1] = trajectory\_distance(lat[1], lon[1],lat[2], lon[2]) $
			\State $t[1] = time[1]$
		
			\For{$i=2$ to $n$} 
				\State $d[i] = trajectory\_distance(lat[i-1], lon[i-1],lat[i], lon[i]) $
				\State $t[i] = time[i] - time[i-1]$
			\EndFor
	
			\State $d_c[] = t_c[] = \phi $ \label{op:cumulative} \Comment{array of commulative distance and delay time for gps points}		
			\State $d_c[1] = t_c[1] = 0 $ 
										
			\For{$i=2$ to $n$} 
				\State $d_c[i] = d_c[i-1] + d[i] $
				\State $t_c[i] = t_c[i-1] + time[i]$
			\EndFor
			
			\State $cl = perform\_clustering(d_c, t_c)$ \Comment{return cluster array}

			\State $verify\_cluster(d_c,t_c,cl)$ \Comment{visual verification, if acceptable then continue to next operation \ref{op:t.test} else move to  operation \ref{op:0} }
			
			\State $df = ensemble(df, d_c, t_c)$
			\State $q = classify\_by\_bonferroni(df)$ \label{op:t.test}  \Comment{return four level of time delay} %\cite{Oliphant2007-pw}
		
		\EndFor
		
		\EndFunction

\end{algorithmic}
\end{algorithm}
%-------------------------------------------------------
%-------------------------------------------------------
\begin{figure}[H]
	\centering
	\includegraphics[width=0.489\textwidth]{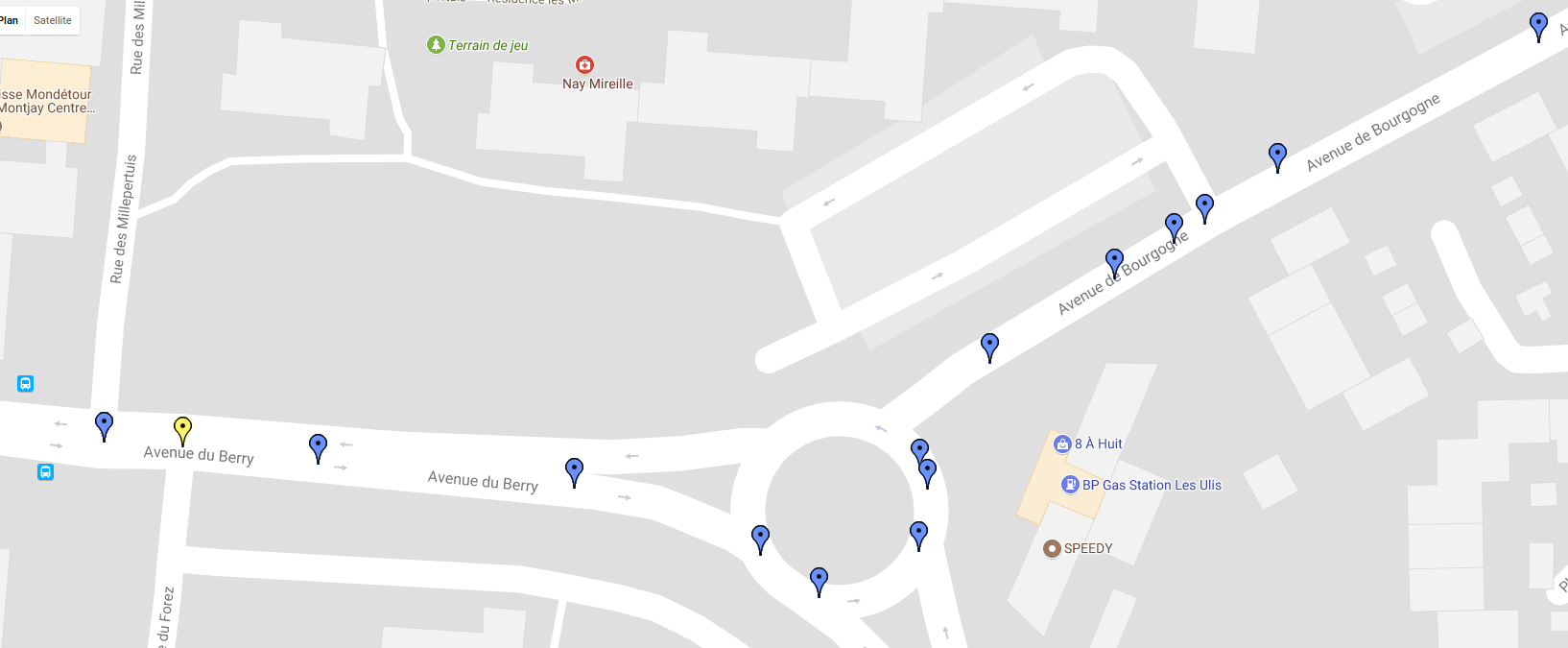}
	\caption{Sorting and aligning of the GPS coordinates. (After)}
	\label{fig:sorting.alignment.after}
\end{figure}
%-------------------------------------------------------
\begin{algorithm}
	\caption{Classification by Bonferroni t test}
	\label{alg:bonferroni}
	\begin{algorithmic}[1]
		\Function{classify\_by\_bonferroni}{$df$}\Comment{where df - data frame with theoratical and real time at each gps point}

		\State $df = do\_t.test(df.ox\_id,\: df.temp\_theo,\: df.temp\_real)$ \Comment{return pVal in df}
		\State $b[][] = multiple\_test(df.pVal, method=bonferroni)$

		\State $t_1 = \overset{b[0]==True}{b[1].min()}$
		\State $t_2 = \overset{b[0]==True}{b[1].max()}$
		\State $t_m = (t_2 - t_1)/2$
		\State $_{t1}^{tm}\textrm{df.class}=$ strong significant
		\State $_{tm}^{t2}\textrm{df.class}=$ significant
		
		\State $f_1 = \overset{b[0]==False}{b[1].min()}$
		\State $f_2 = \overset{b[0]==False}{b[1].max()}$
		\State $f_m = (f_2 - f_1)/2$
		\State $_{f1}^{fm}\textrm{df.class}=$ weak evidence
		\State $_{fm}^{f2}\textrm{df.class}=$ little or no evidence
		
		\State $return\:df.class$ 		
		
		\EndFunction

\end{algorithmic}
\end{algorithm}

%-------------------------------------------------------
The operation \ref{op:ox.1} in algorith \ref{alg:mapping} and \ref{op:ox.2} in algorith \ref{alg:delay.points.identification} was performed by means of OpenStreetMap APIs (OSMnx) graph modeling \cite{Boeing2017-yq}. The OSMnx modeling assists us in finding the nearest point available on the road. It also assists us in finding the distance and the trajectory between two points. If the distance between two points is returned very large then it means the gps coordinates need to be shredded. The challenge is to find the correct route on circular path as well as crossing path as shown by the figure \ref{fig:sorting.alignment.before} and \ref{fig:sorting.alignment.after}. In this situation, a bus route passes by a circle and then cross the same point two times but in the opposite direction. This challenge was also tracked by means of detection of forward and backward points by graphical modeling of the route. If we observe the overall problem from a scientific perspective, the whole of the problem (as shown in the algorithm \ref{alg:delay.points.identification}) can be simplified into two sub-problems as follow:
\begin{itemize}
	\item Cluster the $n$ number of points into $m$ number of clusters.
	\item Perform statistics on each cluster to identify whether the delay in this cluster is significant or not. (see algorithm \ref{alg:bonferroni}) 
\end{itemize}
While formulating the cluster, we have available attributes including place names, original gps coordinates, distance between two points, time required to move between two consecutive points. The time was originally available for a complete trajectory. However, our purpose is to analyze the route over large number of days for a given range of time. The bus produces the gps coordinates randomly from every nine to twenty seconds where in occasionally its frequency was set to ten seconds. Once the bus is passed by a gps coordinate, it is not guaranteed that the bus will pass by this point again in its next trajectory. This means we need to cluster the nearby points. But here a question arises, what are the parameters for the cluster. We must clarify the reason for why we have used the cumulative distance as an input parameter for clustering. We know that k-means is based on euclidean distance. If we use the point to point distance or point to point time then cluster will be composed of random points from the trajectory. The nearest distance points will fall under same cluster. Such a cluster will have no practical value for interpretation. A continuous function is a good candidate for clustering. For this purpose, we derive a new feature of cumulative sum of the distance (see operation \ref{op:cumulative} in algorithm \ref{alg:delay.points.identification}). As the bus starts from its route, its distance naturally increases from the starting point until the end point. This continuous function is also marked by sharp or minor additive value to increase the inter-cluster and decrease the intra-cluster distance.
\par
Another problem in clustering was incoherent alignment of points in the route as shown in the figure \ref{fig:sorting.alignment.before} and \ref{fig:sorting.alignment.after}. For this purpose, we use the graphical modeling of OSMnx \cite{Boeing2017-yq}. The graphical modeling of OSMnx assists us in alignment of the points. Moreover, it is a known fact that selecting appropriate value of $k$ is a scientific problem for which techniques like Silhouette score \cite{Hamerly2002-dr, Rousseeuw1987-zi} and BIC score \cite{Hamerly2002-dr} are well known computational techniques to find optimized value of $k$. We adopted both of them in our methodology. After many experiments, we noticed that in overwhelming situations, BIC \cite{Hamerly2002-dr} was more helpful in finding the exact value of $k$ in k-means clustering. We used the k-mean clustering from SciPy library \cite{Hamerly2002-dr}. 
%-------------------------------------------------------
%	\caption{Rejeu course: Sample of trajectory of a bus covering all of the stations in Orsay sector {\^I}le-de-France}
%	\label{tbl:rejeu.course}
%	\centering
%	\begin{tabular}{|c|c|c|c|c|c|c|c|c|}
%		\hline
%		\makecell{Course\\Numero\\Parc} & \makecell{Course\\Date\\Exploit} & \makecell{Course\\Matricule\\Agent} & \makecell{Course\\MNemo\\Ligne} & \makecell{Course\\Range} & \makecell{Course\\Sens} & \makecell{Course\\NoChainage} & \makecell{Course\\No\\Arret} &
%		\makecell{Course\\Libelle\\Arret} & \makecell{Course\\Horodatage} &
%		\makecell{Course\\Timestamp} \\ \hline
%		3 & 3 & - & - & - & - & - & - & - & - & - \\ \hline
%-------------------------------------------------------
%-------------------------------------------------------
%-------------------------------------------------------
\begin{figure*}
	\includegraphics[width=\textwidth]{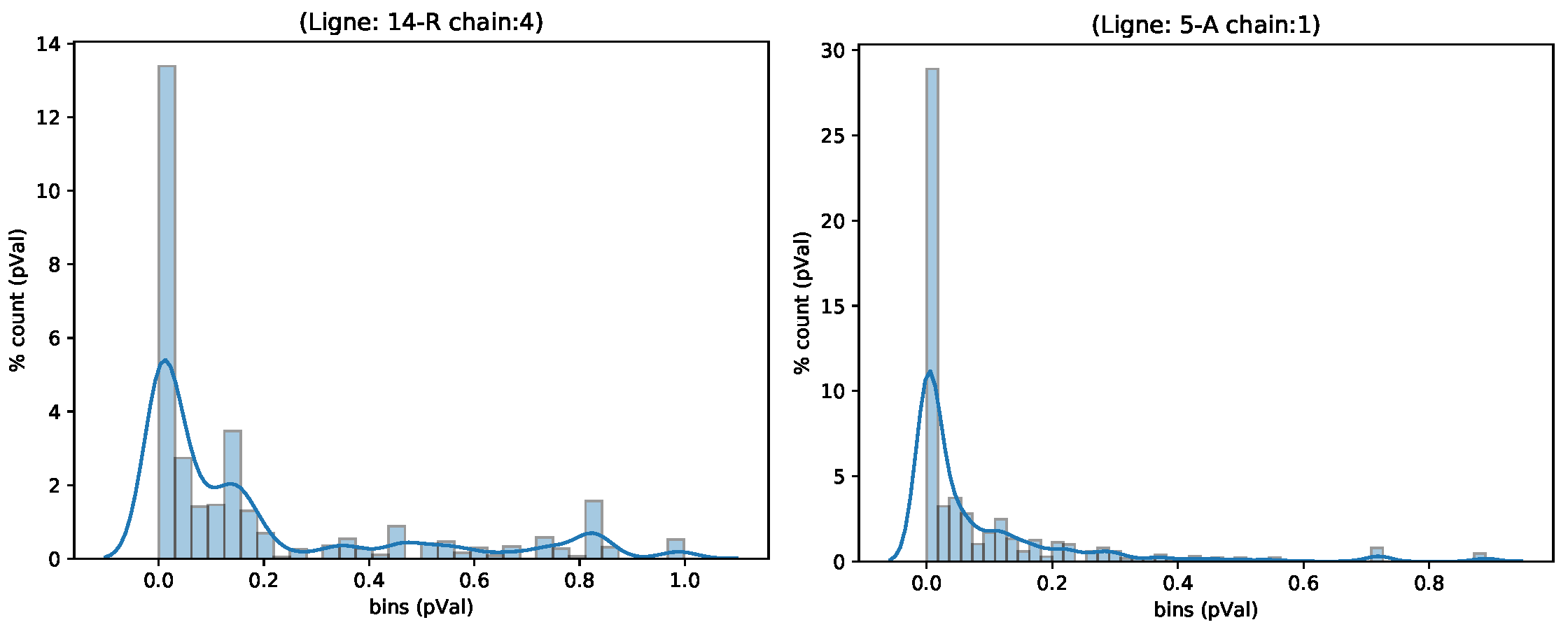}
	\caption{Non convention histogram of statistical analysis after FDR}
	\label{fig:histogram}
\end{figure*}
%-------------------------------------------------------
%%%%%%%%%%%%%%%%%%%%%%%%%%%%%%%%%%%%%%%%%%%%%%%%%%%%%%%%%%%%%%%%%%%%%%%
\section{Results and Discussion}
The empirical validation carried out in this study is independent of any assumption of particular distribution (normal or non normal distribution). However, we know that the student t test is a parametric test.  It means we are supposed to assume that input in the t test follows a normal distribution. The samples with size greater than thirty are resilient in t test as even if there is a violation of normal distribution, it does not cause much problem in the result. Hence the shape of the data does not effect in case of large sample size as the central limit theorem also support the same assumption. However, it is important to ascertain to observe the degree of deviation from the normality of the underlying data. For this purpose, we used the Shapiro-Wilk test. If the null hypothesis (assumption of data normality) is rejected in Shapiro-Wilk test then we have adopted the wilcoxon test as a non parametric test.
\par
\begin{figure*}[!t]
	\includegraphics[width=\textwidth]{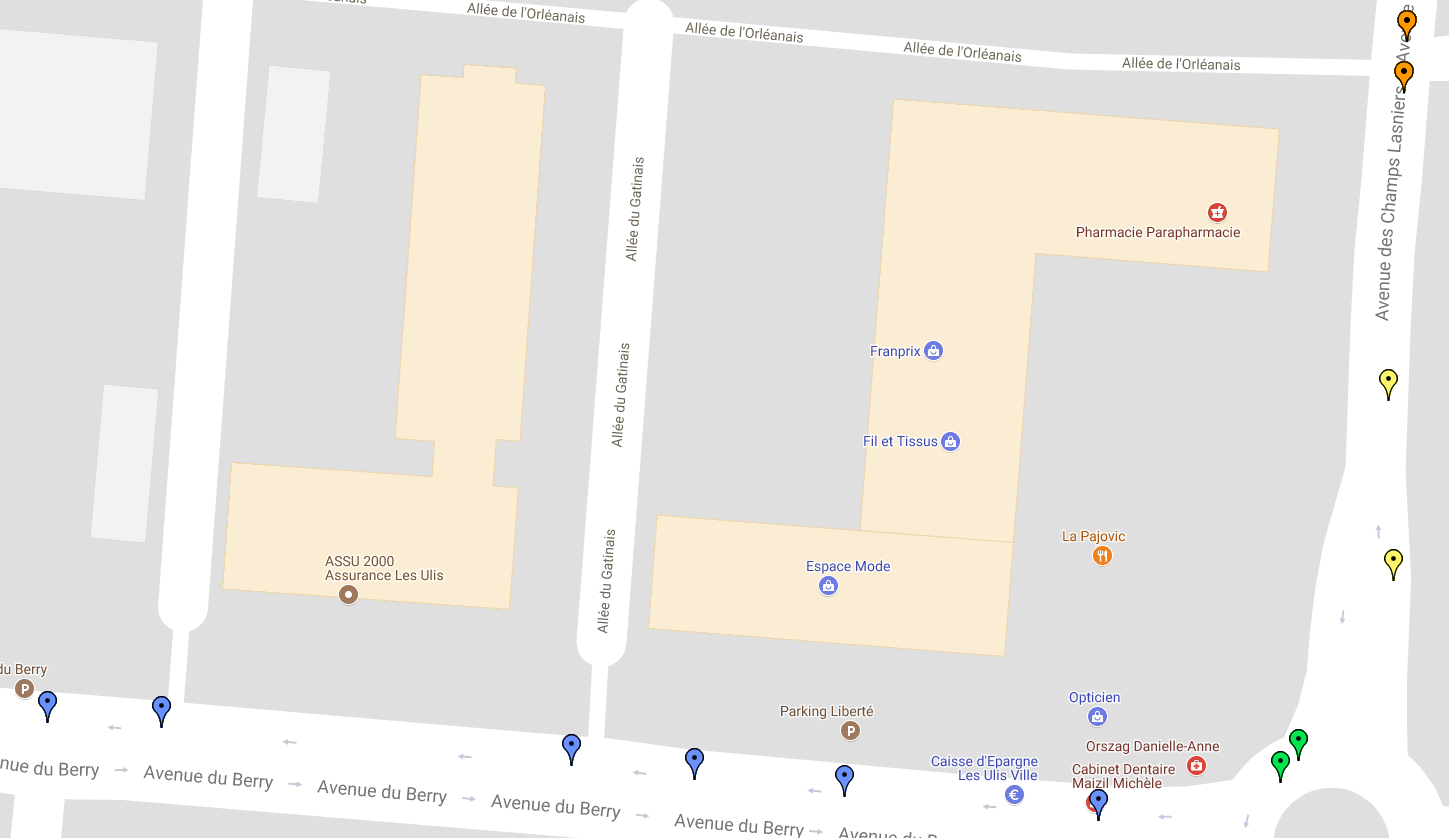}
	\caption{Marking of the points of delay after t test statistics. case 1}
	\label{fig:analysis.1}
\end{figure*}
\begin{figure*}
	\includegraphics[width=\textwidth]{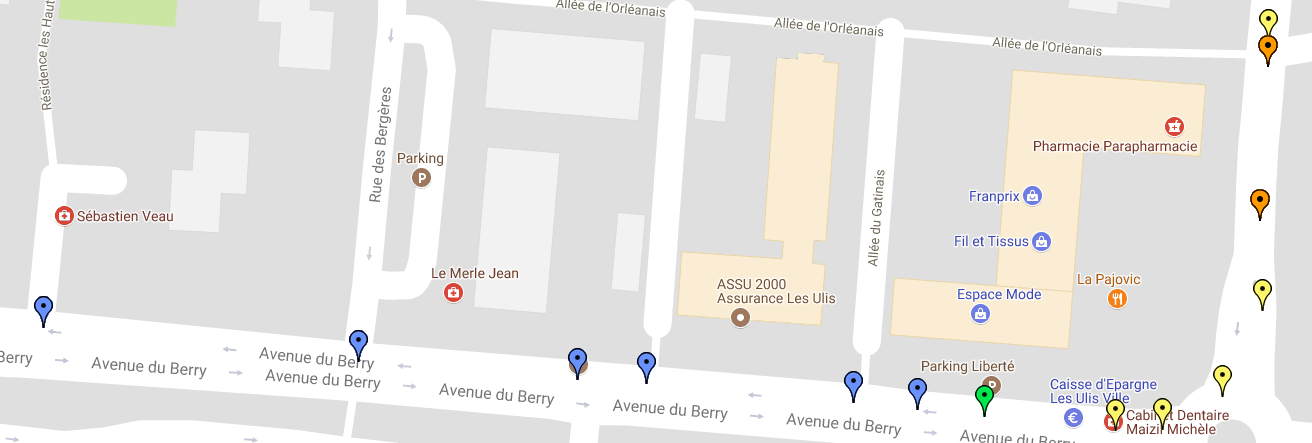}
	\caption{Marking of the points of delay after t test statistics. case 2}
	\label{fig:analysis.2}
\end{figure*}
\begin{figure*}
	\includegraphics[width=\textwidth]{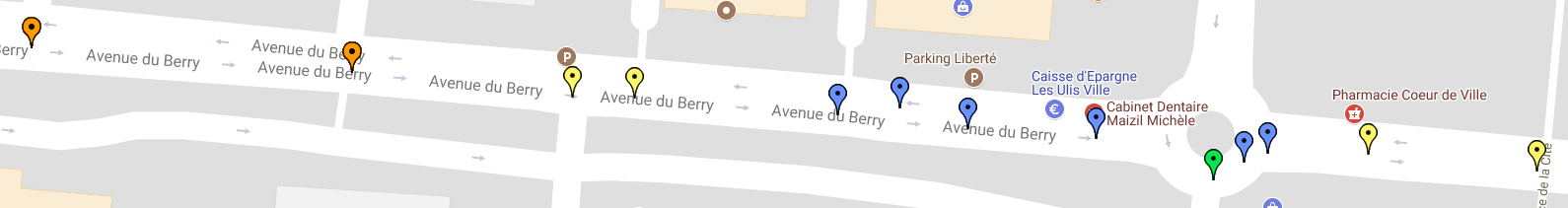}
	\caption{Marking of the points of delay after t test statistics. case 3}
	\label{fig:analysis.3}
\end{figure*}
\begin{figure*}
	\includegraphics[width=\textwidth]{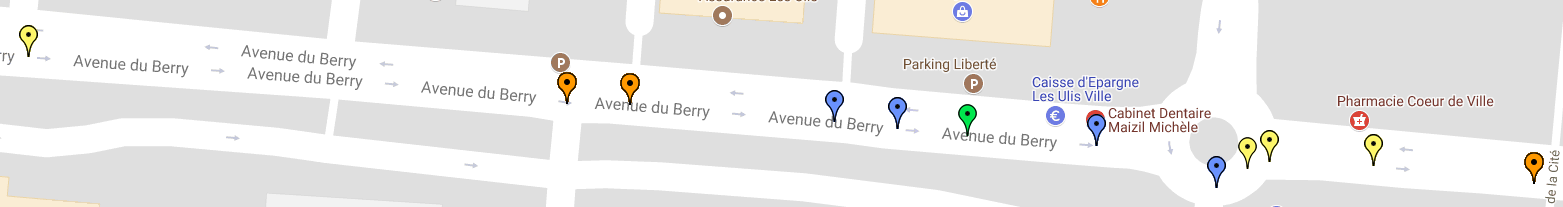}
	\caption{Marking of the points of delay after t test statistics. case 4}
	\label{fig:analysis.4}
\end{figure*}
\begin{figure}[!t]
	\centering
	\includegraphics[width=0.25\textwidth]{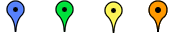}
	\caption{Legends used in Figures \ref{fig:analysis.1},\ref{fig:analysis.2},\ref{fig:analysis.3} and \ref{fig:analysis.4}\\
		\textbf{blue}: strong significant, \textbf{green}: significant\\ \textbf{yellow}: weak evidence, \textbf{orange}: little or no evidence}
	\label{fig:legends}
\end{figure}

We were provided with two different dataset as shown in the figure \ref{fig:rejeu.course} and \ref{fig:rejeu.loc}. The first dataset is "rejeu course" which provides information about a complete route with all of the station name included and the timestamp when the bus pass by. The first dataset is only limited to those courses in which we have the station name and time at which bus passes by. The second dataset contains GPS locations for every line. A total of 19 buses were documented that had usable data, representing an estimated 13,169 trajectories in 92 consecutive days starting from March 2016 to June 2016. Furthermore, each bus has two directions and many sub lines as well. So the number of unique bus routes accumulates to 97. We have mapped both of the dataset as the second dataset holds a composite key comprising of parc number and date. However this composite key is not sufficient as the relationship between both dataset is many to many instead of one to many. Based on this composite key, it was imperative that the second dataset must contain the line, sub line, range and chaine number apart from parc id and date. Keeping in view of this situation, a possible solution was to map the closest time. However it raises another problem of mapping same line but different sub line and same bus line in opposite direction at the same time and date as shown in the figure \ref{fig:sorting.alignment.before} and \ref{fig:sorting.alignment.after}. The steps 7 and 8 in the algorithm were used to handle this problem. The shrewd reader of the article can notice from the figures \ref{fig:analysis.1}, \ref{fig:analysis.2}, \ref{fig:analysis.3} and \ref{fig:analysis.4} that there are different points of interest marked by four different icon colors. The color variation ranges from high significance to low significance. This can assist the viewer in further investigating for why a number of trajectory across many months produce unusual delay at some specific points. A statistical analysis was launched to make a differentiation in the delays at various points as shown by the colors and their explanation in the legend in the figure \ref{fig:legends}.
\par
The following explanation applies for only those cases where the instances exhibit Gaussian distribution behaviour explicitly. In case the data was not following the Gaussian distribution then we adopted non parametric wilcoxon test. We shall explain how we analyze for normal distribution using student t test. We know that the null hypothesis (H\textsubscript{0}) states that there is nothing significant between two parameters. These two parameters include the known fact or existential belief and given parameter to be tested. The p-value acts as a pivot to examine the significance between the known belief and the provided input. Smaller the p-value, the stronger is the evidence against null hypothesis (H\textsubscript{0}). If the p-value is less than 0.01, then we have strong evidence, this strength reduces as we start moving away from 0.01 to a higher value. In general, statisticians have divided the significance level into four bins: very strong from 0.0 to 0.01, strong from 0.01 to 0.05, weak evidence from 0.05 to 0.10 and beyond this there is little or no evidence against H\textsubscript{0}. We performed student t-test to examine about the significance of point delay as a few of the output are shown in the figures \ref{fig:analysis.1}, \ref{fig:analysis.2}, \ref{fig:analysis.3} and \ref{fig:analysis.4}. These figures are explaining the p-value by means of histogram. This is a set of well behaved anti conservative histogram. That flat distribution along the bottom in the figure is null p-values. This provides a uniform distribution. One can observe that they are uniformly distributed between zero and one. This is expected as original definition of p-value implies the same. Simplifying, we can state that under the null, it has a five percent chance for being smaller than 0.05 and so on. The peak which gets closer to zero emanates into alternative hypothesis H\textsubscript{1}. The other aspect is also noteworthy that this H\textsubscript{1} may also be carrying false positive as well. It means, we cannot blindly accept that all values less than 0.05 are significant enough to reject null hypothesis. This requires to perform correction in p-value. We performed multiple testing correction including Bonferroni adjustment and False Discovery Rate ($FDR$) using the guidelines provided in multiple testing correction \cite{Noble2009-pw}. While performing both of the tests, we noticed that Bonferroni adjustment yields no significant match when alpha was set at 0.05. However, FDR estimation method has been reported as sufficient enough for resulting estimates to be provably conservative with respect to a specified H\textsubscript{0} \cite{Noble2009-pw}. The same was noticed in our study.
%-------------------------------------------------------
%-------------------------------------------------------
%-------------------------------------------------------
%-------------------------------------------------------
%-------------------------------------------------------
%---------------------------------------------------------------------
\subsection{Example}
We took statistical analysis for four bus routes. All of these routes are different to each other but they share a segment in common as shown in the figures \ref{fig:analysis.1}, \ref{fig:analysis.2}, \ref{fig:analysis.3} and \ref{fig:analysis.4}. First two routes (top to bottom) are in direction from right to left while the two bottom routes runs in opposite direction (left to right). These are two parallel roads each of them is one-way and there is a clear boundary or limits between both of the ways/roads. If we analyze the maps in the aforementioned  figures, one can notice that before the roundabout, there is a week or minor evidence to reject the null hypothesis H\textsubscript{0}. We shall remind that the null hypothesis is established between two delay vectors, the actual delay and theoretical delay. The hypothesis implies that there is no significant difference between both of these vectors. However upto roundabout (in the figures \ref{fig:analysis.1} and \ref{fig:analysis.2}), we found trivial evidence in favour of rejecting null hypothesis H\textsubscript{0}. Hence we accept into null hypothesis upto these geometry points. However, afterwards, the situation is leading to produce smaller p-value, so we are persuaded in rejecting null hypothesis which means there is a significant time difference in theoretical and real delays. An alternative hypothesis H\textsubscript{1} will be acceptable, hence short to say an unusual or unexpected delay is observed beyond the roundabout. A possible explanation is that, after the roundabout, there are some public offices like bank, mini commercial market and medical centre. The delay might be possibly caused due to the slow passenger movement (getting down or getting into the bus), or due to a crowd of passengers. However, factually there is no bus stop exactly at this point. One key observation is about the busy medical centre whose parking also lies in its vicinity. Although, parking is organized but traffic rush of plenty of vehicles may lead to hamper the speed of traffic flow at the same time slot. This statistical analysis has been drawn for the buses which start around 16:00 o'clock. Certainly these are the peek working hours for public and private offices. The same analysis was performed in other hour-slots and the results were different. Furthermore, the results of the last two routes in the figures \ref{fig:analysis.3} and \ref{fig:analysis.4} logically supports the first two routes in figures \ref{fig:analysis.1} and \ref{fig:analysis.2}. 
In fact, once the bus delay time is affected by external factors, its ripple impact lasts for distance afterwards to some extent. But here, we shall remind our shrewd readers that we are talking about the absolute delay at any point and not the relative delay. Hence the single point is responsible for the statistical analysis taken for as many days as possible. In other words, we are not concerned with the ripple effect of the delay at all. One important observation is that although some of the segments of the points in all of the four bus routes share the same blue color icons except its afterwards or earlier coordinates. In two routes, the whole of the segment is suffering from point delay. Both of these routes are in same direction. The two lower bus routes have shown a part of this segment with colors other than blue. This is obvious that at some timings, only one way of the road is more busy. Hence, the example sufficiently proves that our proposed methodology $DTQS$ is effective enough to provide a point delay across many months using trajectory data of various bus routes.

\begin{table}[!t]
	\caption{Point delay(\%) vs. other delay(\%) of buses in one sector of {\^I}le-de-France}
	\label{tbl:comparison}
	\centering
	\begin{tabular}{|c|c|c|c|c|}
		\hline
		\makecell{Bus \\ id} & observations & \makecell{point \\ delay} & \makecell{stop time \\ delay}  & others \\ \hline
		1 & 4142 & 21 & 65 & 14 \\ \hline
		17 & 18707 & 26 & 58 & 16 \\ \hline
		23 & 15633 & 23 & 67 & 10 \\ \hline
		7 & 87219 & 24 & 67 & 09 \\ \hline
		16 & 1476 & 25 & 65 & 10 \\ \hline
		14 & 37308 & 23 & 66 & 11 \\ \hline
		8 & 18798 & 26 & 62 & 12 \\ \hline
		4 & 62321 & 24 & 62 & 14 \\ \hline
		10 & 94787 & 26 & 61 & 13 \\ \hline
		5 & 61186 & 25 & 65 & 10 \\ \hline
		19 & 67195 & 22 & 67 & 11 \\ \hline
		25 & 116231 & 21 & 67 & 12 \\ \hline
		15 & 91086 & 20 & 68 & 12 \\ \hline
		24 & 128582 & 25 & 65 & 10 \\ \hline
		2 & 391373 & 25 & 64 & 11 \\ \hline
		3 & 299737 & 21 & 65 & 14 \\ \hline
		\multicolumn{2}{|c|}{95\% conf. inter. range} & 22.57 - 24.56 & 63.31 - 65.94 & 10.86 - 12.76 \\ \hline
	\end{tabular}
\end{table}
We took the samples of 20 observations out of total available observations for each bus id as shown in table \ref{tbl:comparison}. Each of the the 20 samples were visually observed and verified. The interesting observations is that point delay usually accounts for the delay from 22.57\% to 24.56\% in 1.96 x standard deviation (96\% confidence interval) across all of the bus lines. Moreover, still it can be noticed that stop time delay is technically the predominant delay but identification of the point delay is also important as it shares approximately one quarter of the total delay as well.
%%%%%%%%%%%%%%%%%%%%%%%%%%%%%%%%%%%%%%%%%%%%%%%%%%%%%%%%%%%%%%%%%%%%%%%
\section{Conclusion}
In this study, we have discussed the quality metric associated with the urban transport network. The delay or early arrival plays significant role in planning the time table of the buses across different seasons. The delay itself is composed of numerous components. There are different types of delays available amongst of which the delay at a specific point has not been much appreciated in the scientific community. There is an increasing impetus in transport operators for in depth investigation of the overall delay. In fact, finding this feature is an analysis for one step ahead of typical timing analysis. We, in this study, have proposed a methodology : Delay Time and Quality of Service ($DTQS$)) which explains for how we can pinpoint those specific paths and geographic coordinates with high importance in terms of major cause of delay. This methodology faces numerous challenges including mapping of incoherent and incomplete datasets drawn out by different vendors. Moreover, we also neglect out the effect of bus stop delays. The most pertinent previous work was related to intersection delays providing quite broaden range of results. However, we have broadened the idea of intersection delay into a concept of geometric point delay with more specific range of outcomes. The point delay accumulates approximately 23.56\% of total overall delay. Moreover, another contribution in this study is that it is not tied to any particular distribution of data. The proposed technique ($DTQS$)) automatically examines the distribution of the data and then selects the parametric or non parametric test accordingly. For the future extension, we propose to estimate the passengers leaving out the bus at any stop so this flow can also be accounted towards the explanation of the overall and point wise delay.
%%%%%%%%%%%%%%%%%%%%%%%%%%%%%%%%%%%%%%%%%%%%%%%%%%%%%%%%%%%%%%%%%%%%%%%

\bibliography{manuscript}
\end{document}